\documentclass[prb, aps, twocolumn, groupedaddress, superscriptaddress, nofootinbib]{revtex4-2}
\usepackage[dvipsnames]{xcolor}             %
\usepackage[colorlinks=true,linkcolor=Blue,citecolor=Blue,urlcolor=Blue]{hyperref}
\usepackage{graphicx}
\usepackage{bbm}
\usepackage{bm}
\usepackage{amsmath}
\usepackage{amssymb}
\usepackage{mathptmx}
\usepackage{hhline}
\usepackage[version=3]{mhchem}
\usepackage{array}
\usepackage[normalem]{ulem}
\usepackage{multirow}

\usepackage{tikz}
\usepackage{enumitem}

\DeclareMathAlphabet{\mathcalCM}{OMS}{cmsy}{m}{n}
\newcommand{\bn}{\ensuremath{\mathcalCM{N}}}

\newcommand{\Jxy}{J_{xy}}
\newcommand{\Jz}{J_{z}}
\newcommand{\hc}{h_{c2}}
\newcommand{\Tnem}{T_{\mathrm{nem}}}

\newcommand{\Tbkt}{T_{\mathrm{BKT}}}

\graphicspath{{figures/}}

\begin{document}
\title{Ground state and thermal phase diagrams of the magnetized \\spin-$1/2$ XXZ antiferromagnet on the triangular lattice} 

\author{Owan \surname{Cho}}
\altaffiliation{These authors contributed equally to this work.}
\affiliation{Department of Physics, Korea Advanced Institute of Science and Technology, Daejeon 34023, Korea}

\author{Seungho \surname{Lee}}
\altaffiliation{These authors contributed equally to this work.}
\affiliation{Department of Physics, Sungkyunkwan University, Suwon 16419, Korea}
\affiliation{Institute of Basic Science, Sungkyunkwan University, Suwon 16419, Korea}

\author{Hyun-Yong \surname{Lee}}
\email{hyunyong@korea.ac.kr}
\affiliation{Division of Semiconductor Physics, Korea University, Sejong 30019, Korea}
\affiliation{Department of Applied Physics, Graduate School, Korea University, Sejong 30019, Korea}

\author{SungBin \surname{Lee}}
\email{sungbin@kaist.ac.kr}
\affiliation{Department of Physics, Korea Advanced Institute of Science and Technology, Daejeon 34023, Korea}

\begin{abstract}
The spin-1/2 XXZ antiferromagnet on the triangular lattice in a magnetic field is a paradigmatic frustrated quantum magnet, yet its ground-state phase diagram at generic anisotropy and its thermal phase structure remain incompletely understood. Here, using infinite density-matrix renormalization group, infinite projected entangled-pair operator, and exponential tensor renormalization group methods, supported by exact diagonalization, we resolve both issues. At zero temperature, we map the phase diagram continuously from the Ising limit to the Heisenberg point and find that the nonclassical $\Psi$ supersolid is a stable phase just below saturation at all anisotropies studied: it borders the one-third-magnetization up-up-down plateau directly in the strongly easy-axis regime and the semiclassical $V$ phase toward the Heisenberg point. 
In the strongly easy-axis regime, we uncover an intrinsically finite-temperature bond-nematic phase above the zero-temperature plateau--$\Psi$ boundary, in which the three bond directions become spontaneously inequivalent without transverse spin coherence or three-sublattice order. A hard-core-boson mapping reveals its origin: once thermal fluctuations destroy phase coherence, the field-doped holes settle on domain walls of the local up-up-down order, which are stabilized by quantum kink delocalization and aligned by its $Z_3$ registry. Our results call for reexamining high-field states in triangular-lattice cobaltates such as Na$_2$BaCo(PO$_4$)$_2$ and K$_2$Co(SeO$_3$)$_2$.
\end{abstract}

\date{\today}
\maketitle

Frustrated spin-$1/2$ magnets on the triangular lattice provide a paradigmatic setting in which quantum fluctuations, geometric frustration, and magnetic fields conspire to generate unconventional phases of matter \cite{Starykh15}. A prime example is the triangular-lattice XXZ antiferromagnet in a magnetic field, which hosts a one-third-magnetization up-up-down (UUD) plateau bordered by spin-supersolid phases with broken spin-$U(1)$ and three-sublattice translational symmetries at zero temperature~\cite{Miyashita86,NishimoriMiyashita86,ChubukovGolosov91,Yamamoto14,PhysRevB.91.081104}. Finite-temperature studies have also examined thermal ordering in the classical XXZ model and its Heisenberg limit~\cite{MiyashitaKawamura85,KawamuraMiyashita85,Miyashita86,StephanSouthern00,Seabra11}, as well as thermodynamics and spin correlations in the spin-$1/2$ model~\cite{Elstner93,Chen19,gao2022spin,Ulaga24}. Nevertheless, the thermal phase structure of the high-field quantum supersolid remains incompletely understood, particularly in the strongly easy-axis regime. An important question is whether heating merely restores the symmetries broken in the supersolid or can instead stabilize an intermediate phase with a distinct form of order.

Here we combine infinite density-matrix renormalization group (iDMRG), infinite projected entangled-pair operator (iPEPO), and exponential tensor renormalization group (XTRG) calculations to determine the ground-state phase diagram and how the high-field supersolid melts at finite temperature. At zero temperature, we establish the phase diagram continuously from the strong easy-axis regime to the Heisenberg point, showing that the $\Psi$ supersolid forms a stable phase between the $V$ and fully polarized phases throughout this range. At finite temperature, we uncover an intermediate bond-nematic phase above the high-field supersolid. In this regime, transverse magnetic coherence and three-sublattice positional order are absent, whereas the bond correlations spontaneously distinguish the three lattice directions. This orientational order is distinct from the sublattice-domain order of the low-temperature supersolid. A hard-core-boson mapping and domain-wall analysis suggest a microscopic origin. Excess magnetization above the UUD plateau is accommodated at zero temperature by a coherent condensate of doped holes; once this transverse coherence is thermally lost, the same holes are more favorably accommodated by domain-wall configurations stabilized through quantum kink delocalization. 
Chiral wall wetting may further facilitate composite UUD domains, as an unfavorable wall can split into two favored walls separated by an intermediate domain if this lowers the interfacial free energy. The proliferation of such domains could disorder the global $Z_3$ UUD registry while remaining compatible with collective wall alignment, offering a possible route to the intermediate bond-nematic regime.

\emph{Model and ground-state phase diagram.}---
We study the spin-$1/2$ XXZ antiferromagnet on the triangular lattice,
\begin{equation}
H=
\sum_{\langle ij\rangle}
\left[
J_{xy}\left(S_i^xS_j^x+S_i^yS_j^y\right)
+J_zS_i^zS_j^z
\right]
-h\sum_i S_i^z,
\label{eq:hamiltonian}
\end{equation}
where $\langle ij\rangle$ denotes nearest-neighbor pairs,
$J_z>0$, and $J_{xy}/J_z\in[0,1]$ interpolates
from the Ising limit to the Heisenberg point. We determine the
ground-state phase diagram using iDMRG on infinite cylinders
\cite{White92,Ian08,zauner-stauber18prb,SMnote}.

\begin{figure}[t]
    \centering
    \includegraphics[width=0.8\columnwidth]
    {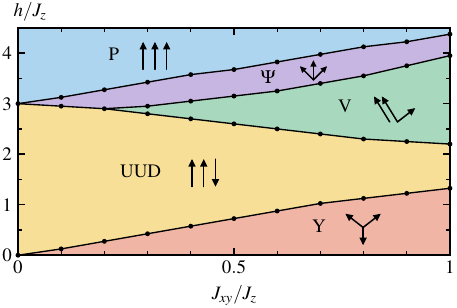}
    \caption{Ground-state phase diagram of Eq.~\eqref{eq:hamiltonian} in the $(J_{xy}/J_z,h/J_z)$ plane obtained by iDMRG. Black circles mark the estimated phase boundaries, and solid lines are guides to the eye. Arrows sketch the spin orientations on the three sublattices; P denotes the fully polarized phase.}
    \label{fig:gs}
\end{figure}

Fig.~\ref{fig:gs} shows the resulting phase diagram. The one-third-magnetization UUD plateau persists from the Ising limit to the Heisenberg point~\cite{Miyashita86,ChubukovGolosov91}. 
In the strong easy-axis regime, increasing the field yields the $Y$ supersolid, UUD plateau, $\Psi$ supersolid, and fully polarized phase (P), while a $V$ supersolid intervenes between the UUD and $\Psi$ phases toward the Heisenberg point.
The $\Psi$ phase occupies a finite field interval between the $V$ and P phases.

Previous studies generally assigned the high-field coplanar state to the conventional $V$ phase using semiclassical spin-wave theory near the Heisenberg limit, cluster mean-field calculations with cluster-size scaling on the easy-axis side, and finite-cylinder DMRG at $J_{xy}/J_z\simeq0.60$~\cite{ChubukovGolosov91,Yamamoto14,gao2022spin}. Near saturation, however, exact diagonalization in fixed-magnon sectors does not distinguish $V$ from $\Psi$ in the thermodynamic limit~\cite{Yamamoto17}. Recent iPEPS calculations for K$_2$Co(SeO$_3$)$_2$, with $J_{xy}/J_z\approx0.07$, found the high-field state to be $V$ only at bond dimension $D=1$ and $\Psi$ for $D\geq2$, and reported the same for $J_{xy}/J_z\lesssim0.4$ at $h=3J_z$~\cite{XuHasik25}. Our iDMRG results agree with these iPEPS results in the strong easy-axis regime and show that the $\Psi$ phase extends toward the Heisenberg point.

\begin{figure}[tb]
\includegraphics[width=\columnwidth]{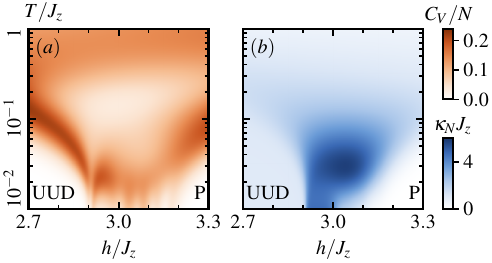}
\caption{Exact diagonalization (ED) of Eq.~\eqref{eq:hamiltonian} at $J_{xy}/J_z=0.1$ on a 21-site torus that preserves the threefold lattice rotation $C_3$ and accommodates three-sublattice order.
(a) Specific heat per site, $c_V$, and (b) susceptibility $\kappa_\bn$ of the bond-nematic order parameter to a $C_3$-breaking bond anisotropy, versus $h/J_z$ and $T/J_z$ (logarithmic $T$ axis).
UUD and P label the plateau and fully polarized ground states of the cluster.}
\label{fig:ed}
\end{figure}

\textit{Exact-diagonalization signatures of bond nematicity.---}
We next examine the finite-temperature bond-nematic response above the UUD plateau using exact diagonalization of Eq.~\eqref{eq:hamiltonian} at $\Jxy/\Jz=0.1$ on a 21-spin torus compatible with both $C_3$ rotation and three-sublattice order~\cite{SMnote}. The bond-nematic order parameter is defined as
\begin{equation}
\bn=\sum_{\alpha=0}^{2}\omega^{\alpha}B_{\alpha},
\qquad
\omega=e^{2\pi i/3},
\label{eq:nematic}
\end{equation}
where $B_\alpha=2\langle S^x_iS^x_j+S^y_iS^y_j\rangle_{\langle ij\rangle\parallel\mathbf{e}_\alpha}$, with the average taken over all bonds parallel to the lattice direction $\mathbf{e}_\alpha$. A nonzero $|\bn|$ signals spontaneous breaking of the lattice $C_3$ rotational symmetry. Here, $C_3$ denotes the site-centered lattice rotation that cyclically permutes the three bond directions. The order parameter $\bn$ is translationally invariant and therefore distinguishes bond-orientational order from the $Z_3$ domain choice of the three-sublattice state. In particular, a three-sublattice spin texture does not by itself imply $\bn\neq0$: each bond direction contains $AB$, $BC$, and $CA$ bonds in equal number, so that any three-sublattice state, including the $V$ and $\Psi$ supersolids, has $\bn=0$. Bond nematicity would represent an additional symmetry breaking beyond three-sublattice magnetic order.

The symmetric thermal state
has $\langle \bn \rangle=0$, so we probe its susceptibility
$\kappa_\bn=\partial\langle\mathrm{Re}\,\bn\rangle/\partial\eta|_{\eta=0}$,
with the source coupled as $H(\eta)=H-\eta N_s\mathrm{Re}\,\bn$,
where $N_s=21$ is the number of spin sites.
The $C_3$-symmetric torus does not impose a preferred bond direction,
but finite-size effects remain.
Figure~\ref{fig:ed}(b) shows an enhanced bond-nematic susceptibility
at magnetic fields above the UUD phase, relative to the response
within the plateau.
On the 21-spin torus, this field range extends from the
zero-temperature upper edge of the UUD plateau, $h_{c2}^{(21)}$,
to the saturation field, $h_{\rm sat}$.
At $h=3J_z$, $\kappa_\bn$ has a broad maximum near $T=0.03J_z$
within the temperature range shown and remains enhanced upon
further cooling.

\begin{figure*}[t]
    \centering
    \includegraphics[width=2\columnwidth]{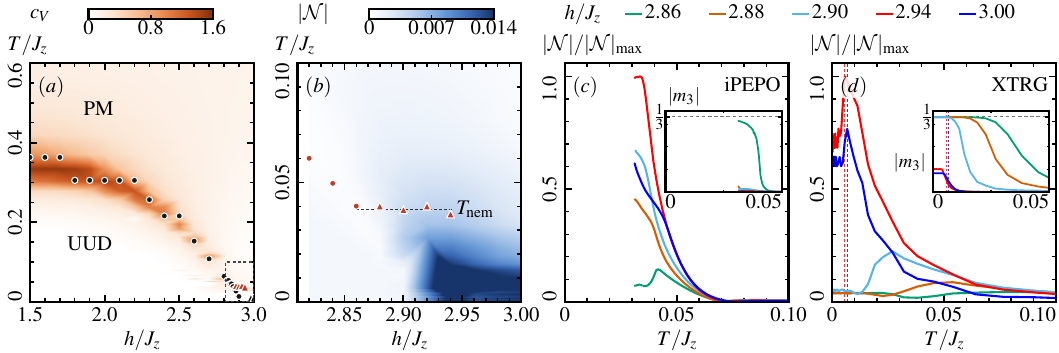}
\caption{Finite-temperature specific heat and bond nematicity at $J_{xy}/J_z=0.1$. (a) iPEPO specific heat $c_V$, calculated with bond dimension $D=14$ and environment dimension $\chi_{\mathrm{CTM}}=28$, supplemented by $D=24$, $\chi_{\mathrm{CTM}}=36$ data near the high-field edge of the UUD plateau. Black circles and triangles indicate the estimated UUD transition temperatures and additional high-field specific-heat peaks obtained from XTRG, respectively. Orange triangles in the boxed region mark the high-field specific-heat anomalies obtained from iPEPO. (b) XTRG bond-nematic amplitude $|\bn|$ in the boxed region of (a). Orange circles and triangles indicate the positions of iPEPO specific-heat peaks, and the dashed line marks the estimated $\Tnem$. (c,d) Temperature dependence of the normalized bond-nematic amplitude $|\bn|/|\bn|_{\max}$ from (c) iPEPO and (d) XTRG at the fields specified in the shared legend. The insets show the corresponding three-sublattice order parameter $|m_3|$, with horizontal dashed lines marking $|m_3|=1/3$.}
    \label{fig:tn}
\end{figure*}

The specific heat provides additional context. At $h=3\Jz$, it has a broad $\Jz$-scale maximum and a lower-temperature feature of order $\Jxy$ [Fig.~\ref{fig:ed}(a)], consistent with a separation of interaction and kinetic energy scales. The enhanced bond-nematic response thus provides suggestive evidence for a bond-nematic phase above the UUD plateau. Finite-cluster susceptibility, however, cannot establish spontaneous $C_3$ symmetry breaking, the extent of the phase, or its transition temperatures; we therefore turn to thermodynamic-limit methods.


\emph{Finite-temperature results: a bond-nematic phase.}---
To determine whether this enhanced response develops into spontaneous order, we use iPEPO calculations directly in the
thermodynamic limit~\cite{Czarnik12,Kshetrimayum19}, supplemented
by XTRG calculations on cylinders~\cite{ChenXTRG18,SMnote}. The iPEPO uses a $3\times3$ unit cell of the triangular lattice, compatible with three-sublattice order, and is cooled from $T=\infty$ by symmetric Trotter steps using the simple-update scheme~\cite{motoyama22, motoyama25}. We focus on $\Jxy/\Jz=0.1$ at fields close to the boundary between the UUD plateau and the high-field supersolid, $\hc$. Fig.~\ref{fig:tn}(a) presents the resulting thermal phase diagram. 
The specific-heat peaks trace the apparent phase boundary between the UUD and paramagnetic phases, while weak additional anomalies appear at lower temperatures near $h_{c2}$, orange triangles in Fig.~\ref{fig:tn}(a) and (b); see the SM for an enlarged view~\cite{SMnote}.
These anomalies coincide with the steepest rise of the iPEPO bond-nematic amplitude [Fig.~\ref{fig:tn}(c)]: upon cooling from the paramagnet, the three bond orientations become inequivalent. We identify the $c_V$ maximum with $\Tnem$ [$\simeq0.036$--$0.039\Jz$ for $2.88\le h/\Jz\le2.94$]. The onset appears continuous. Since the broken symmetry is the threefold rotation $C_3$, such a transition is expected to belong to the two-dimensional three-state Potts class~\cite{Wu82}.

Figs.~\ref{fig:tn}(c) and \ref{fig:tn}(d) show the field and temperature dependence of the bond nematicity obtained by iPEPO and XTRG, respectively. Both methods find the largest bond-nematic amplitude at $h/\Jz=2.94$ among the fields and temperature ranges shown, with a pronounced enhancement relative to the UUD plateau and fully polarized regimes. In iPEPO, the signal persists upon increasing the bond dimension $D$ and environment dimension $\chi_{\rm CTM}$~\cite{SMnote}.

The two approaches access complementary parts of the thermal evolution. 
The iPEPO calculations capture the onset of bond nematicity at $\Tnem$ and its growth upon cooling, while the three-sublattice order parameter $|m_3|$ remains suppressed in this regime [inset of Fig.~\ref{fig:tn}(c)]. Here,
$
m_3=\frac{1}{3}\sum_{\alpha=0}^{2}\omega^{\alpha}\langle S^z_{\alpha}\rangle,
~~ \omega=e^{2\pi i/3},
$ 
with $\alpha=0,1,2$ labeling the three sublattices.
 XTRG extends to lower temperatures and reveals a nonmonotonic evolution: at $h/\Jz=2.94$ and $3.00$, $|\bn|$ reaches a maximum and then decreases upon further cooling. Within the intermediate-temperature window where the bond nematicity is growing, $|m_3|$ does not show a detectable ordering signal [insets of Fig.~\ref{fig:tn}(d)]. The growth of $|m_3|$ occurs at lower temperatures, where the bond-nematic response is reduced.

Taken together, the two-dimensional iPEPO results and complementary XTRG data support an intermediate bond-nematic phase with broken lattice $C_3$ symmetry. This orientational order is distinct from the three-sublattice order of the low-temperature supersolid and does not simply reflect its spin texture. Because the cylinder geometry explicitly distinguishes lattice directions, XTRG probes the orientational response but does not independently establish spontaneous $C_3$ symmetry breaking. The low-temperature evolution observed in XTRG is consistent with leaving the nematic regime, but does not locate the lower phase boundary in two dimensions.


\emph{Microscopic domain-wall mechanism.}---
To understand the bond-nematic order, we use the exact
hard-core-boson mapping~\cite{MatsubaraMatsuda56},
$b_i^\dagger=S_i^-$ and $n_i=\frac12-S_i^z$, which transforms
Eq.~\eqref{eq:hamiltonian} into
\begin{equation}
H_{\rm b}
=
t\sum_{\langle ij\rangle}
\left(b_i^\dagger b_j+\mathrm{h.c.}\right)
+V\sum_{\langle ij\rangle}n_i n_j
-\mu\sum_i n_i+\mathrm{const.},
\label{eq:boson}
\end{equation}
with $t=J_{xy}/2$, $V=J_z$, and $\mu=3J_z-h$. Magnetizing beyond the UUD plateau thus corresponds to hole doping of the one-third-filled boson crystal~\cite{PhysRevB.79.020409}. For $V/t\gg1$, the nearest-neighbor repulsion acts as an approximate blockade, within which doped holes can lower their energy by forming domain walls of the three-sublattice order, each carrying one hole per three rows~\cite{SMnote}. On the spatially anisotropic triangular lattice, such quantum domain walls stabilize an incommensurate domain walls on supersolid whose orientation is fixed by the exchange anisotropy~\cite{Zhang16}; here the lattice is isotropic, and the orientation must be selected spontaneously.

\begin{figure}[t]
    \centering
    \includegraphics[width=1\columnwidth]
    {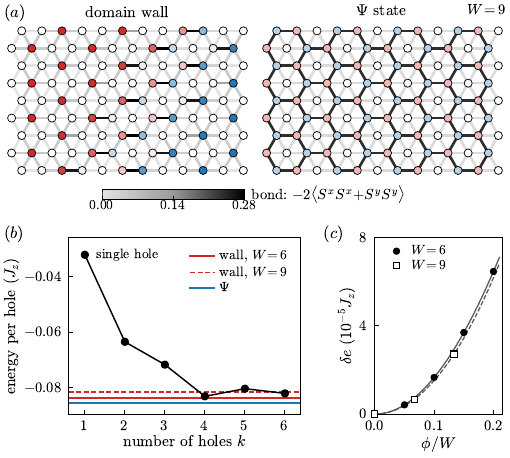}
   \caption{Domain walls at $\Jxy/\Jz=0.1$. (a) iDMRG on a cylinder of circumference $W=9$: a domain wall (left) and the uniform $\Psi$ state (right). Each site is colored by its sublattice (red or blue) and drawn darker the more it is occupied by a boson; empty sites are white. Bond shading gives $-2\langle S^x_iS^x_j+S^y_iS^y_j\rangle$ [cf.\ Eq.~\eqref{eq:nematic}], so dark bonds mark bosons resonating between two sites. (b) Kinetic energy per hole for $k$ holes on a finite $W=6$ cylinder, compared with an isolated domain wall ($W=6$, solid; $W=9$, dashed) and the $\Psi$ state. (c) Flux-induced energy of $\Psi$ against the phase gradient $\phi/W$; its curvature gives the stiffness $\rho_s$. See SM~\cite{SMnote}.}
    \label{fig:dw}
\end{figure}

The walls are stabilized by their kink dynamics. From row to row, the position of a wall shifts by one of two allowed displacements, and a kink is a place where consecutive displacements differ. Exchanging two adjacent displacements requires a single boson hop between exactly degenerate states (Fig.~S3 of the SM~\cite{SMnote}), so kinks propagate with the bare amplitude $t$ and map onto a spin-1/2 XX chain. These hops appear directly in iDMRG: Fig.~\ref{fig:dw}(a) shows a domain wall on a cylinder of circumference $W=9$, held in place by weak pinning fields, whose kinks resonate on bonds of a single direction, next to the uniform $\Psi$ state. The wall free energy per row is
\begin{equation}
f_w(T,\mu)=\frac{\mu}{3}-\frac{T}{2\pi}\int_{-\pi}^{\pi}\ln\!\left[2\cosh\frac{t\cos k}{T}\right]dk,
\label{eq:fw}
\end{equation}
which reduces to the zero-point gain $-2t/\pi$ for $T\ll t$ and to the configurational entropy $-T\ln 2$ for $T\gg t$. The wall gain is thus not degraded by heating: it retains its zero-point value up to $T\sim t$ and crosses over to an entropic gain beyond. Localized alternatives cannot match this gain: isolated holes have no first-order mobility~\cite{SMnote}. On a finite cylinder, DMRG finds that doped holes bind: the energy per hole falls steeply as holes are added and approaches the energy of an isolated domain wall once the holes suffice to form domain walls [Fig.~\ref{fig:dw}(b)]. ED on the 21-spin torus is consistent with this: the three-hole sector selected at the plateau edge gains more energy per hole than an isolated hole, by an amount comparable to the domain-wall estimate. This finite cluster cannot accommodate a uniform condensate, so the comparison establishes the ordering of energy scales rather than the ground state~\cite{SMnote}.

At $T=0$, iDMRG nevertheless finds the uniform $\Psi$ supersolid of Fig.~\ref{fig:gs} above $\hc$: the doped holes condense, with a gain that relies on two-dimensional phase coherence and is lost at a Berezinskii--Kosterlitz--Thouless (BKT) transition at $\Tbkt\approx(\pi/2)\rho_s\approx0.006J_z$, estimated from the phase stiffness $\rho_s$ of $\Psi$ [Fig.~\ref{fig:dw}(c)], well below $\Tnem$~\cite{SMnote}. The energy difference is small: on cylinders, $\Psi$ lies only slightly below an isolated domain wall [Fig.~\ref{fig:dw}(b)]. Because this energy difference is a small fraction of $t$, whereas the free-energy gain of a wall grows with temperature through its kink entropy [Eq.~\eqref{eq:fw}], thermal wall fluctuations may overcome it already at temperatures well below $t$.

The wall's local $O(t)$ resonances can persist above $\Tbkt$ without global two-dimensional phase coherence. At a fixed wall direction, hopping favors one chirality. An unfavorable wall can split into two favored walls separated by an intermediate UUD domain if this lowers the free energy per projected length, providing a route to chiral wall wetting~\cite{HuseFisher1984,FisherWetting1984,SMnote}. We propose that wall proliferation disorders the global UUD registry, while collective wall alignment retains a preferred bond axis without long-range periodic wall order.
Further heating restores $C_3$ at $\Tnem$ through an apparently continuous transition compatible with three-state Potts criticality~\cite{SMnote,Wu82}. The decrease of $|\bn|$ and growth of $|m_3|$ on cooling [Fig.~\ref{fig:tn}(d)] support an intermediate-temperature nematic, although the cylinder data do not locate its lower boundary in two dimensions. 

Balancing the $O(t)$ kinetic gain, with $t\propto\Jxy$, against the classical wall tension suggests an $O(\Jxy)$ field scale near $h=3\Jz$, consistent with the iDMRG energy comparisons~\cite{SMnote}. The thermal window depends additionally on competing free energies and wall interactions. The clock sum defining $\bn$ cancels the $C_3$-symmetric UUD background and, in the dilute-wall limit, measures wall alignment weighted by segment length and local bond anisotropy. For the untilted walls in Fig.~\ref{fig:dw}(a), resonant hopping is concentrated on the bond direction perpendicular to the wall, whereas the corresponding correlations in $\Psi$ extend over all three directions~\cite{SMnote}.



\emph{Discussion and outlook.}---
Using iPEPO and XTRG calculations supported by exact diagonalization, we have shown that the high-field supersolid of the triangular-lattice XXZ antiferromagnet melts through an intermediate bond-nematic phase with neither transverse coherence nor three-sublattice positional order. 
Our hard-core-boson analysis suggests that the bond-nematic phase originates from the orientational ordering of quantum-stabilized domain walls.
The nematic is therefore not rotational order inherited from the supersolid, whose three-sublattice texture is itself $C_3$ symmetric, but a distinct thermal phase of the wall ensemble. This picture reconciles a $C_3$-symmetric $\Psi$ ground state with a $C_3$-broken phase above it: heating removes the two-dimensional coherence that favors the condensate while leaving the one-dimensional kink gain of the walls intact.

At zero temperature, the $\Psi$ supersolid intervenes between the $V$ and fully polarized phases from the strong easy-axis regime to the Heisenberg point, connecting limits previously addressed by separate approaches~\cite{Miyashita86,Yamamoto14,PhysRevB.91.081104,Yamamoto17,gallegos2024phase,Ulaga25,Kadosawa26}. The high-field state of magnetized triangular-lattice cobaltates~\cite{Shirata12,Susuki13,gao2022spin,sheng22pnas,xiang2024giant,Zhu25}, usually identified with the $V$ phase, should accordingly be reexamined close to saturation.

K$_2$Co(SeO$_3$)$_2$, with $J_{xy}/J_z\approx0.07$~\cite{Zhu25}, has an anisotropy close to that studied here and is a natural candidate for the proposed bond nematic. Neutron scattering could search for directional anisotropy in diffuse magnetic correlations, while specific-heat measurements could help identify the relevant thermal scales. The lower boundary of the nematic phase in two dimensions and the sequence of three-sublattice and transverse ordering upon cooling remain unresolved, as does the extent of the nematic regime toward the Heisenberg point, where the blockade description becomes less controlled. How further-neighbor and additional anisotropic exchanges modify wall interactions and affect the competition between the nematic and supersolid phases also remains an open question.

\emph{Acknowledgments}---We would like to thank Sasha Chernyshev for helpful comments and suggestions. H.-Y.L. was supported by the Basic Science Research Program through the National Research Foundation (NRF) of Korea funded by the Ministry of Science and ICT (MSIT) (Grant Nos. RS-2023-00220471 and RS-2025-16064392). S.B.L and O.C were supported by National Research Foundation of Korea (RS-2026-25475824, RS-2023-00281839) and Air Force Office of Scientific Research under FA23862514054. SL was supported by the National Research Foundation of Korea (NRF) grant funded by the Korea government(MSIT) (Grant No. 2023R1A2C1002644)

\bibliography{references}

\end{document}


\title{Supplementary Materials} 
\date{\today}
\maketitle

\section{Numerical details}
\subsection{iDMRG}

We determine the ground-state phase diagram using infinite
density-matrix renormalization group (iDMRG) calculations on
infinitely long cylinders with periodic boundary conditions
around the circumference. We primarily use cylinders of
circumference $W=9$ and a maximum matrix-product-state (MPS) bond dimension
$\chi=1600$, and compare selected results with $W=6$.
Both circumferences are compatible with three-sublattice order.

\begin{figure}[t]
    \centering
    \includegraphics[width=0.99\columnwidth]{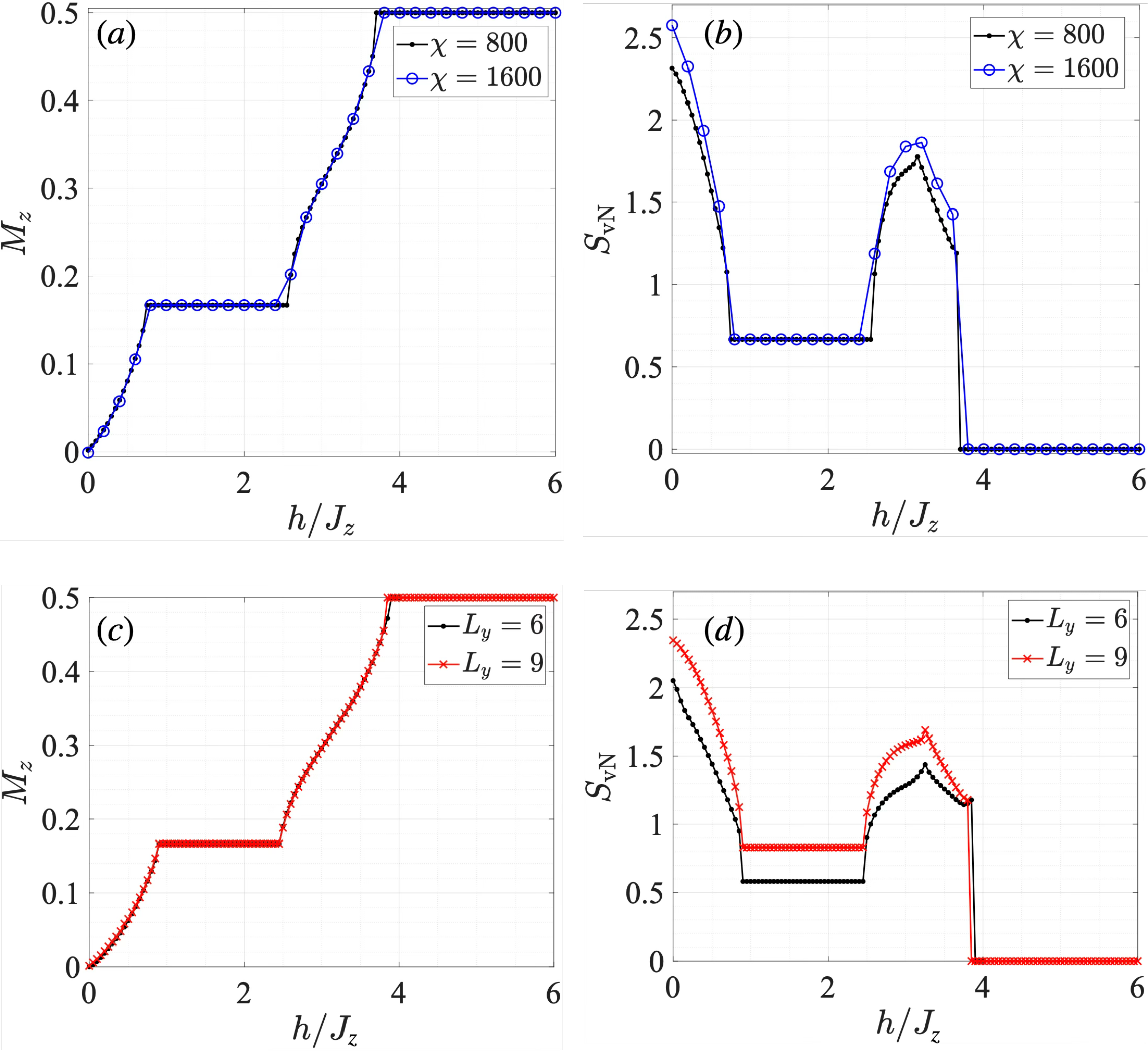}
\caption{
Bond-dimension and circumference dependence of the iDMRG
results. (a,b) Magnetization per site $M_z$ and von Neumann
entanglement entropy $S_{\rm vN}$ at $J_{xy}/J_z=0.5$ and the circumference
$L_y=9$, for MPS bond dimensions $\chi=800$ and $1600$.
(c,d) The same observables at $J_{xy}/J_z=0.6$ for cylinder
circumferences $L_y=6$ and $9$.
The magnetic field is expressed in units of $J_z$.
The magnetization varies smoothly across the $V$--$\Psi$
boundary, whose location is estimated from the kink in
$S_{\rm vN}$.}
    \label{fig:sm_dmrg}
\end{figure}

Figures~\ref{fig:sm_dmrg}(a,b) compare the field dependence of
the magnetization per site $M_z$ and the von Neumann
entanglement entropy $S_{\rm vN}$ for $\chi=800$ and $1600$
at $W=9$. The magnetization curves nearly coincide,
including the UUD plateau at $M_z=1/6$ and the fully polarized
region at $M_z=1/2$. The entropy retains a visible $\chi$
dependence in the supersolid regions, while showing similar
features as a function of field. In the fully polarized
region, $S_{\rm vN}$ vanishes.

The total magnetization varies smoothly across the
$V$--$\Psi$ boundary and therefore does not distinguish the
two supersolids. We estimate this boundary from the kink in
$S_{\rm vN}$ and identify the phases by examining the local
spin expectation values on the three sublattices on either
side. The resulting spin configurations correspond to the
$V$ and $\Psi$ states, respectively. Thus, the entropy kink
provides an estimate of the boundary, while the
three-sublattice spin configurations determine the phase
assignment.

Figures~\ref{fig:sm_dmrg}(c,d) compare cylinders with the circumference
$L_y=6$ and $9$. The magnetization curves are close, and the
entropy shows similar plateau, high-field kink, and saturation
features, although its magnitude depends on the circumference.
For the representative scans shown, the estimated phase
boundaries change little between the two circumferences.

\subsection{iPEPO}
For thermal phase diagrams, we use the infinite projected
entangled pair operator (iPEPO), as implemented in the Tensor Network Solver
(\textsc{TeNeS}) package \cite{motoyama22,motoyama25}.  We employ a \(3\times3\) unit cell,
\((L,W)=(3,3)\).  The thermal density operator \(\rho(\beta)\) is obtained by
imaginary time evolution starting from the infinite-temperature
density operator, \(\rho(0)=I\). Evolution is performed by applying the
symmetric Trotter update
\begin{align}
\rho(\beta+2\tau)=e^{-\tau H}\rho(\beta)e^{-\tau H}.
    \label{eq:Trotter}
\end{align}
For \(J_{xy}/J_z=0.1\), and for each value of \(h\), the
evolution is carried out up to \(\beta=45\) using an iPEPO bond dimension
\(D=14\) and a time step \(\tau=0.015\).  The tensor-network
environment is contracted using the corner transfer matrix renormalization group
(CTMRG) method with a CTM bond dimension \(\chi_{\rm CTM}=28\).

\begin{figure}[t]
    \centering
    \includegraphics[width=1.02\columnwidth]{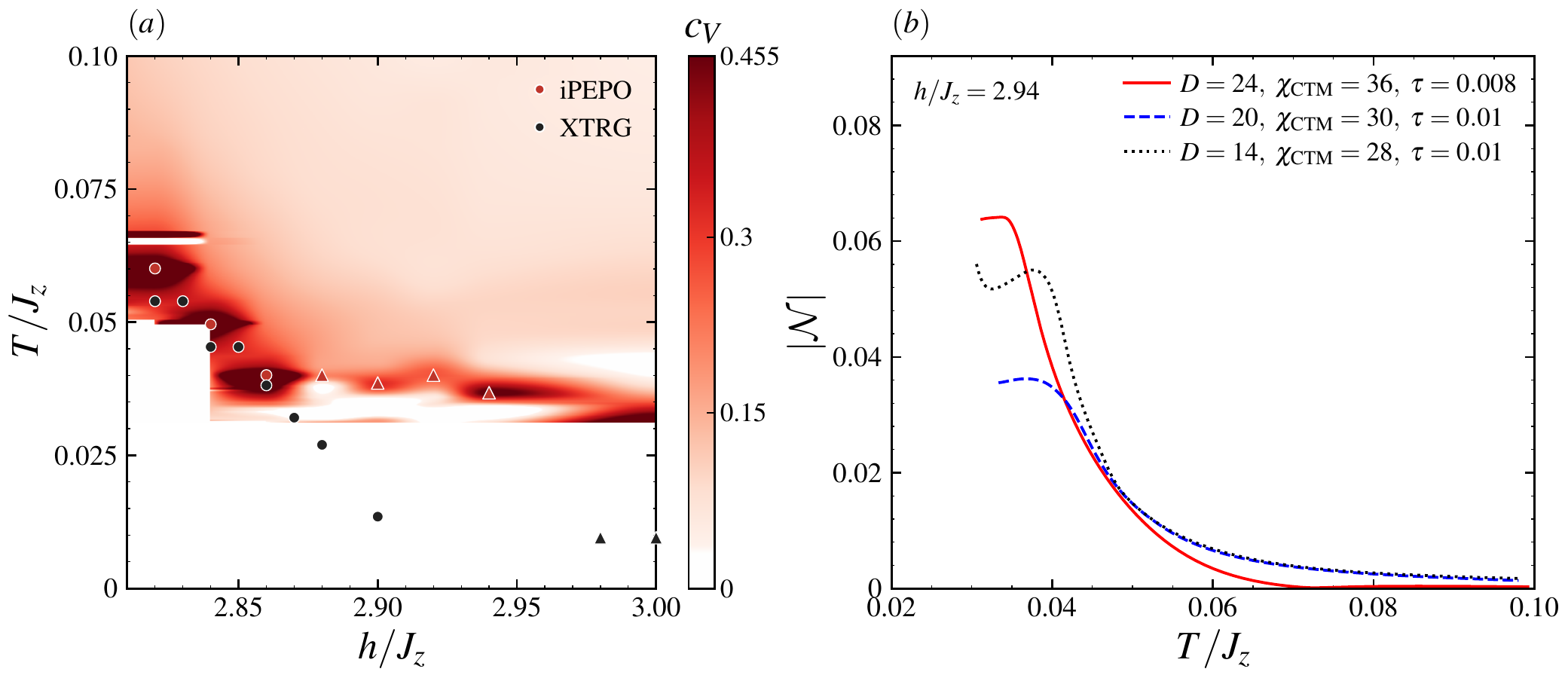}
\caption{Specific heat and bond nematicity near $h_{c_2}$ at $J_{xy}/J_z=0.1$.
(a) High field specific-heat plot obtained from iPEPO calculations
with $D=24$, $\chi_{\rm CTM}=36$, and $\tau=0.008$.
Red and black colors indicate specific-heat peak temperatures
obtained from iPEPO and XTRG, respectively.
Circles mark peaks associated with the paramagnetic--UUD
transition, while triangles denote additional low-temperature
broad peak at higher fields.
(b) Temperature dependence of the bond nematic amplitude
$|\bn|$ at $h/J_z=2.94$ for three iPEPO parameter sets:
$(D,\chi_{\rm CTM},\tau)=(24,36,0.008)$
(solid line), $(20,30,0.01)$ (dashed line), and
$(14,28,0.01)$ (dotted line).
Here, $D$ denotes the bond dimension, $\chi_{\rm CTM}$ denotes the CTMRG environment dimension
and $\tau$ is the imaginary-time evolution step.
}
    \label{fig:sm_cv_bn_comparison}
\end{figure}

Near the $h_{c_2}$, we additionally perform calculations with higher bond dimension,
$D=20,24$, $\chi_{\rm CTM}=30,36$, and $\tau=0.01,0.008$, respectively.
Figure~\ref{fig:sm_cv_bn_comparison}(a) shows the specific-heat over the field range
$2.8 < h/J_z \le 3.00$ based on $D=24$, $\chi_{\rm CTM}=36$, and $\tau=0.008$ datasets.
In addition to the specific-heat peak associated with the
paramagnet--UUD transition, additional low temperature broad peaks appear at higher fields. 
In the iPEPO results, these broad peak $T_{\rm nem}$ occur near
$0.036-0.039J_z$ for $h/J_z=2.88-2.94$, on the temperature
scale over which the bond-nematic amplitude grows upon cooling.

Figure~\ref{fig:sm_cv_bn_comparison}(b) examines the dependence of
this nematic response on the numerical parameters at
$h/J_z=2.94$, a field with a pronounced enhancement of nematic amplitude.
All three parameter sets exhibit a substantial increase in
$|\bn|$ upon cooling, supporting the qualitative
robustness of the nematic response across the tested settings.

\subsection{XTRG}
For $J_{xy}/J_z=0.1$, we perform exponential tensor renormalization group (XTRG) calculations on cylinders of circumference $W=6$ and length $L=24$, with periodic boundary conditions around the circumference and open ends. The unnormalized thermal density operator $\rho(\beta)=e^{-\beta H}$ is represented as a matrix product operator with maximum bond dimension $\chi=300$. We initialize $\rho(\beta_0)\simeq I-\beta_0H$ at $\beta_0J_z=10^{-4}$ and iteratively multiply the density operator by itself, \begin{equation} \rho(2\beta)=\rho(\beta)\cdot\rho(\beta), \end{equation} thereby doubling the inverse temperature at each step. At each step, the MPO product is compressed in a variational manner. Four interleaved sequences give a spacing $\Delta\ln\beta=\ln 2/4$. The specific heat is obtained by differentiating the energy per site using a centered five-point stencil on the logarithmic temperature grid.

\subsection{Exact diagonalization}
\label{sec:ed}
We use the tori spanned by $\mathbf{L}_1=(2,2)$, $\mathbf{L}_2=(-2,4)$
(12 spins) and $\mathbf{L}_1=(4,1)$, $\mathbf{L}_2=(-1,5)$ (21 spins), in
units of the triangular-lattice primitive vectors. Both superlattices are
mapped onto themselves by the $120^\circ$ rotation, and the
three-sublattice coloring closes on both. The $C_3$ symmetry makes the
three bond directions equivalent and enforces $\langle\mathrm{Re}\,\bn \rangle=0$
in the unperturbed Gibbs state; the numerical residuals are of order
$10^{-12}$--$10^{-11}$. Three-sublattice commensurability also places the
zone-corner momentum $K$ on the cluster momentum grid, so the
single-magnon saturation field $h_{\rm sat}=3J_z+\tfrac32J_{xy}$ is
represented exactly.

The Hamiltonian is diagonalized fully in every block of total $S^z$ and
lattice momentum. Because $S^z_{\rm tot}$ is conserved, the field enters
only through $E(h)=E(h_0)-(h-h_0)S^z_{\rm tot}$, so one set of spectra at
$h_0=3J_z$ yields the complete $(h,T)$ maps of Fig.~2 of the main text.
The nematic susceptibility is obtained from the exact free energy of
$H(\eta)=H-\eta N_s\,\mathrm{Re}\,\bn$, i.e., $J_{xy}\to J_{xy}-2\eta$
on the bonds along $\mathbf e_0$ and $J_{xy}\to J_{xy}+\eta$ on the other
two, as
$\kappa_{\bn}=(\beta N_s)^{-1}\partial^2\ln Z/\partial\eta^2|_{\eta=0}$.
The second derivative is taken by centered differences at
$\eta/J_z=2\times10^{-4}$, $10^{-4}$, and $5\times10^{-5}$ with two
Richardson steps. On a finite torus $\kappa_{\bn}$ is a response,
not an order parameter, and it cannot establish spontaneous symmetry
breaking.

\begin{table}[tb]
\caption{Ground energy per doped hole relative to the plateau,
$[E(x)-E_{\rm UUD}]/n_h$ in units of $J_{xy}$, at $h=3J_z$ and
$J_{xy}/J_z=0.1$, against the hole doping $x=n_h/(N_s/3)$. The
domain-wall estimate is $-6t/\pi=-(3/\pi)J_{xy}\simeq-0.955J_{xy}$.}
\label{tab:ed}
\begin{ruledtabular}
\begin{tabular}{lcccccc}
$x$ & $1/7$ & $1/4$ & $2/7$ & $3/7$ & $1/2$ & $1$ \\
\hline
12 spins & -- & $-0.561$ & -- & -- & $-1.105$ & $+0.072$ \\
21 spins & $-0.327$ & -- & $-0.708$ & $-0.935$ & -- & $+0.071$ \\
\end{tabular}
\end{ruledtabular}
\end{table}

Table~\ref{tab:ed} lists the ground energy per doped hole at $h=3J_z$.
The high-field edge of the plateau selects the three-hole sector on
21 spins and the two-hole sector on 12 spins, at dopings $x=3/7$ and
$1/2$. On 21 spins the three-hole sector gains $0.935J_{xy}$ per hole,
compared with $0.327J_{xy}$ for an isolated hole and with the domain-wall
estimate $0.955J_{xy}$. Repeating the calculation at
$J_{xy}/J_z=0.05$, $0.02$, and $0.01$ and extrapolating the per-hole
energies to $J_{xy}\to0$ gives a coefficient of $-2.01t$ for three holes,
$-1.22t$ for two, and $0$ for one, i.e., an isolated hole has no
first-order gain. The fully doped entries, $+0.071$--$0.072J_{xy}$,
reflect the virtual-hopping energy $3J_{xy}^2/(4J_z)$ that the plateau
loses per removed boson. These finite clusters cannot accommodate the
uniform condensate, and the energies alone do not identify the
real-space structure of the doped states; they establish the ordering
of energy scales.

The specific heat near $h=3J_z$ shows two separated features. The
high-temperature maximum is set by $J_z$ alone: on 12 spins it does
not move when $J_{xy}/J_z$ is reduced from $0.1$ to $0.01$, and in the
plateau interior it is accompanied by the entropy $\ln3/12$ per spin of
the three UUD states. At $h=3J_z$ the constrained manifold of the $t=0$
limit consists of the 62 configurations without adjacent down spins on
12 spins; its entropy, $\ln62/12=0.344$ per spin, is the intermediate
plateau of the entropy for $t\ll T\ll J_z$ ($0.3437$ at
$J_{xy}/J_z=0.01$ and $T=0.1J_z$). At finite $t$ this entropy is released
at a temperature of order $J_{xy}$, which gives the lower feature.

\section{Domain wall mechanism}

\subsection{Domain wall proliferation}

\begin{figure}[!ht]
 \centering
 \includegraphics[width=\columnwidth]{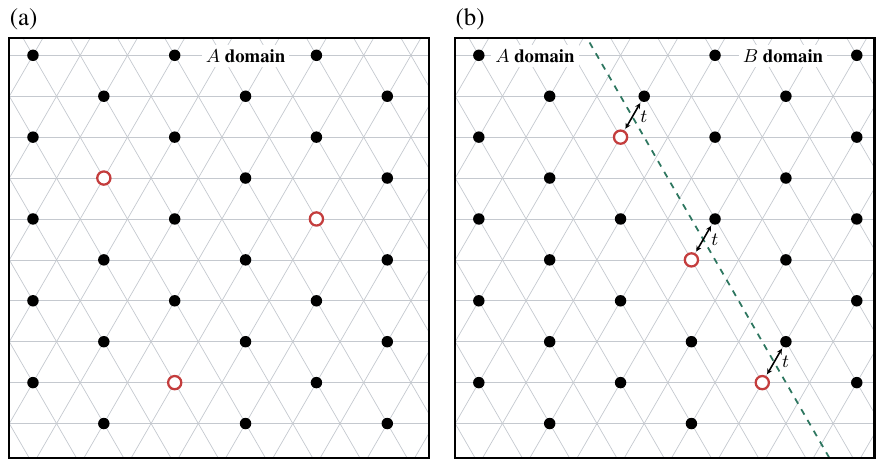}
 \caption{
(a) Isolated holes and (b) a domain-wall configuration at the same density, 
 at the UUD--supersolid phase transition point.
Black and red denote occupied sites (down spins) and holes relative to UUD ordering;
the dashed line marks the wall.
Isolated hole hopping requires second order virtual hopping
through states that violate nearest neighbor exclusion,
yielding an energy gain of $O(t^2/J_z)$ per hole.
In contrast, first order resonant hopping across the wall
(arrows labeled $t$ in (b)) yields $O(t)$. }
 \label{fig:sm_dw_proliferation}
\end{figure}

At zero temperature, the transition between the UUD solid and the
coplanar supersolid at high fields is known to become first order
in the regime of strong easy axis anisotropy,
$J_{xy}/J_z\lesssim0.4$, with $\mu_c=3J_z-h_{c2}>0$
\cite{PhysRevB.91.081104}. A microscopic picture of this
first order transition can be developed by analogy with the
mechanism of domain wall proliferation proposed by Wessel and
Troyer~\cite{WesselTroyer2005}.
At low temperatures satisfying $T,t,|\mu|\ll J_z$, the relevant
Hilbert space is restricted to configurations in which no two
bosons occupy nearest neighbor sites. The corresponding projected
hard hexagon Hamiltonian is
\begin{equation}
    H
    =
    P
    \left[
        t\sum_{\langle ij\rangle}
        \left(b_i^\dagger b_j+b_j^\dagger b_i\right)
        -\mu\sum_i n_i
    \right]
    P
    +O\left(\frac{t^2}{J_z}\right),
\end{equation}
where $P=\prod_{\langle ij\rangle}(1-n_i n_j)$ enforces
$n_i n_j=0$ on every bond between nearest neighbors.

Starting from the UUD solid at $\langle n \rangle=1/3$, hopping of a boson
into an isolated hole requires virtual intermediate states
outside the hard hexagon manifold. The resulting kinetic energy
gain is therefore of order $-t^2/J_z$. By contrast, holes
arranged in a domain wall of the type considered by Wessel and
Troyer allow resonant hopping within the constrained manifold
at first order in $t$ [Fig.~\ref{fig:sm_dw_proliferation}].

This $O(t)$ kinetic-energy gain favors domain-wall configurations over isolated holes. Because the resonant processes do not require global transverse phase coherence, they can remain relevant after coherence is thermally lost. Together with the numerically observed bond anisotropy, this suggests a possible interpretation of the finite-temperature nematic in terms of fluctuating walls with collective orientational order.

\subsection{Domain wall free energy and chirality}

\begin{figure}[t]
    \centering
    \includegraphics[width=0.8\columnwidth]{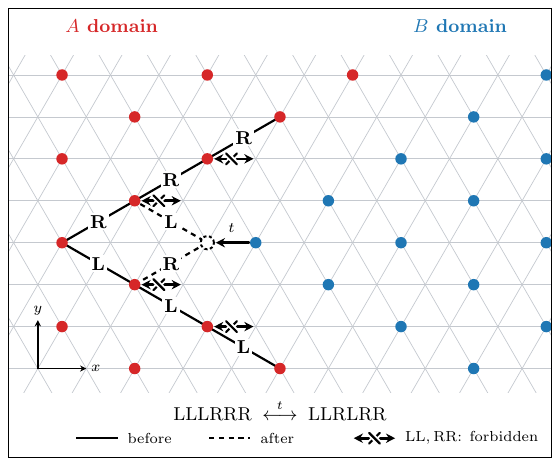}
    \caption{Local move of an $A\!\to\!B$ wall. The last occupied $A$ site
    in successive rows steps by $L$ or $R$. A single boson hop $t$ within
    one row exchanges an adjacent $LR$ pair, e.g.,
    $LLLRRR\leftrightarrow LLRLRR$ (solid: before; dashed: after); the
    pairs $LL$ and $RR$ allow no hopping.}
    \label{fig:LR_flip}
\end{figure}

The three close packed UUD domains are denoted by sublattice $A$, $B$, and
$C$. Consider an $A\!\to\!B$ wall, with $A$ on the left and $B$
on the right, and choose rows along the lattice direction in
which the sites occur in the order $A,B,C$. Three empty sites
separate the occupied domains in each row, compared with two
in the bulk. The wall therefore carries $1/3$ of a hole per
row and costs $\mu/3$ per unit length.

Taking the row spacing as unity, the hard hexagon constraint
allows only two relative positions of the last occupied $A$
site in successive rows, $\Delta{\bf r}_{L,R}=(\mp\sqrt{3},1)$.
Denote these by $L$ and $R$ sites. The chain occupation records
whether a displacement is $\Delta{\bf r}_R$,
\begin{equation}
 |L\rangle\equiv|0\rangle_{R},\qquad
 |R\rangle\equiv|1\rangle_{R} .
 \label{eq:wall_local_states}
\end{equation}
The wall is then an $L/R$ sequence.
The pairs $LL$ and $RR$ do not allow hopping. The only local move is
\begin{equation}
 \langle\cdots LR\cdots|H|\cdots RL\cdots\rangle=t,
 \label{eq:wall_flip}
\end{equation}
together with its reverse [Fig.~\ref{fig:LR_flip}].
The exchange of $L$ and $R$ is a hop of a hard-core particle on a
one-dimensional chain. Writing $c_i^\dagger c_{i+1}\equiv b_R^{\dagger}b_{R+1}$ for
the boson hop between the $R$ and $R+1$ sites in row $i$, the domain wall
is described by the XX-chain Hamiltonian
\begin{equation}
 H_{\rm w}=\frac{\mu l}{3}
   +t\sum_{i=1}^{l}
    \left(c_i^\dagger c_{i+1}+c_{i+1}^\dagger c_i\right),
 \label{eq:wall_xx_chain}
\end{equation}
for length $l$.
The Jordan Wigner transformation gives free fermions with
dispersion $\epsilon_k=2t\cos k$. To fix the average direction,
introduce $\mu_{\rm w}$ to control the occupation
$n_{\rm w}\equiv\langle c_i^\dagger c_i\rangle$.
The free energy per unit length at that occupation is
\begin{align}
 f_{\rm w}(T,\mu,\mu_{\rm w})
 &=\frac{\mu}{3}+\mu_{\rm w}n_{\rm w}\nonumber\\
 &\quad-\frac{T}{2\pi}\int_{-\pi}^{\pi}
       \ln\!\left[1+e^{-(\epsilon_k-\mu_{\rm w})/T}\right]dk,
 \label{eq:wall_free_energy}\\
 n_{\rm w}
 &=\frac{1}{2\pi}\int_{-\pi}^{\pi}
       \frac{dk}{1+e^{(\epsilon_k-\mu_{\rm w})/T}} .
 \label{eq:wall_occupation}
\end{align}

Without a directional constraint, $\mu_{\rm w}=0$ and
$n_{\rm w}=1/2$ at every temperature. Equal numbers of
$L$ and $R$ make the average direction vertical, with
\begin{equation}
 f_{\rm w}(T,\mu)=\frac{\mu}{3}
 -\frac{T}{2\pi}\int_{-\pi}^{\pi}
       \ln\!\left[2\cosh\!\left(\frac{t\cos k}{T}\right)\right]dk .
 \label{eq:wall_unbiased_free_energy}
\end{equation}
At zero temperature,
\begin{equation}
 f_{\rm w}(0,\mu)=\frac{\mu}{3}-\frac{2t}{\pi}.
 \label{eq:wall_zero_temperature}
\end{equation}
Thus the resonance gain is $6t/\pi$ per
hole. Setting this wall energy to zero gives the estimate
$\mu_c\simeq6t/\pi$ for the transition out of the UUD solid.
This estimate omits corrections of order $t^2/J_z$
(Sec.~\ref{sec:dw_dmrg}).

The preferred average direction of a wall also determines
its chirality. Translations cyclically permute $A,B,C$,
so the walls $A\to B$, $B\to C$, and $C\to A$ share the
same chirality, denoted by $[+]$, while their reverses are
denoted by $[-]$. By rotational symmetry, the
preferred average directions of opposite chiralities differ
by $\pi/3$. We take the unbiased $[+]$ wall, with free
energy per unit length $f_{\rm w}(T,\mu)$, to be vertical.
A parallel $[-]$ wall then requires an all-$L$ or all-$R$
sequence in its own row frame, eliminating
$LR\leftrightarrow RL$ resonance and giving a free energy
of $\mu/6$ per unit length.

During melting, the formation of an intermediate domain
lowers the free energy by replacing the $[-]$ wall with
two $[+]$ walls, a process known as domain wall
wetting~\cite{FisherWetting1984}.
Huse and Fisher argued that such proliferation
of walls of the same chirality can be relevant to
commensurate melting~\cite{HuseFisher1984}.
 For the untilted $[+]$ walls considered here, hopping resonance contributes to $B_\alpha$ along the bond direction perpendicular to their mean orientation. Their proliferation could disorder the global UUD registry and, if accompanied by collective orientational order, support the emergence of a bond-nematic phase.

\section{DMRG study of domain walls}
\label{sec:dw_dmrg}

\subsection{Setup}
We examine the ingredients of the wall picture at $J_{xy}/J_z=0.1$
($t=J_{xy}/2=0.05J_z$) with $U(1)$-symmetric DMRG
(TeNPy~\cite{Hauschild18}) on YC cylinders, in which one
nearest-neighbor bond direction runs around the circumference, with
$W=6$ and $9$ sites around it; $\bm a_1$ runs along the cylinder and
$\bm a_2$ around it. In an infinite MPS whose unit cell
holds $L_x$ columns, a UUD domain fits only if $L_x$ is a multiple of
three. For $L_x\not\equiv0\pmod 3$ every cell must contain one wall
($A\!\to\!B$ for $L_x\equiv1$, $A\!\to\!C$ for $L_x\equiv2$), carrying
$W/3$ holes. The resulting wall array is extrapolated linearly in
$1/L_x^2$ to an isolated wall ($L_x=7$, $10$, $13$ for $W=6$; $7$, $8$,
$10$, $11$ for $W=9$). Uniform states are obtained in cells with
$L_x$ a multiple of three at fixed hole number.

For configurations without nearest-neighbor bosons, the Ising
energy exceeds that of the UUD solid by exactly $3J_z$ per hole,
so we use the kinetic energy
\begin{equation}
 q=E-Ne_{\rm UUD}-3J_z n_h ,
 \label{eq:dmrg_q}
\end{equation}
which vanishes without hopping and includes both first-order hops and
second-order virtual processes; $\varepsilon=q/n_h$ denotes the
kinetic energy per hole. In a field, the energy relative to the
plateau is $n_h(3J_z-h)+q$.

The phase stiffness $\rho_s$ of the $\Psi$ state is obtained by threading
a flux $\phi$ through the cylinder, i.e., attaching the phase
$e^{i\phi\Delta y_{ij}/W}$ to every hopping, where $\Delta y_{ij}$ is the
displacement of the bond around the circumference, and fitting
\begin{equation}
 e(\phi)-e(0)=\tfrac12A_{\rm site}\,\rho_s\,(\phi/W)^2+O(\phi^4),
 \label{eq:dmrg_stiffness}
\end{equation}
where $A_{\rm site}=\sqrt3/2$ is the area per site; the UUD solid shows no
response. $T_{\rm BKT}\approx(\pi/2)\rho_s$ is then an estimate from the
ground-state stiffness of finite-width cylinders, not a bound.

\subsection{Hole binding}
\begin{figure}[t]
\centering
\includegraphics[width=\columnwidth]{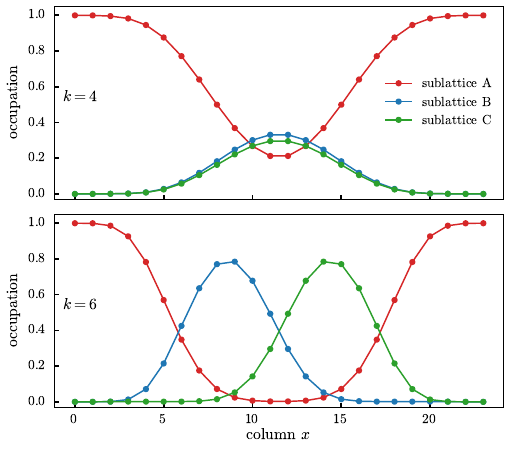}
\caption{Sublattice occupation, averaged over each column, along the
finite $W=6$ cylinder for the lowest states with $k=4$ and $k=6$ holes.
The ends are pinned to the $A$ solid. Four holes deplete $A$ in the
middle in favor of $B$ and $C$ (two walls, $A\!\to\!B\!\to\!A$ and
$A\!\to\!C\!\to\!A$ in superposition); six holes form three walls,
$A\!\to\!B\!\to\!C\!\to\!A$.}
\label{fig:dw_binding}
\end{figure}

Finite DMRG on a $W=6$ cylinder of 24 columns, whose ends are pinned to
the $A$ solid, gives the lowest state with $k$ holes. Since the length
is a multiple of three and both ends are $A$, this geometry does not
force any wall: walls can only appear in pairs ($A\!\to\!B\!\to\!A$ or
$A\!\to\!C\!\to\!A$, $2W/3=4$ holes) or triples
($A\!\to\!B\!\to\!C\!\to\!A$, $W=6$ holes). For $k\le5$, clustered and
spread initial states converge to the same state; for $k=6$, a start
with three walls reaches a lower energy, which we show. A single hole
gains only at second order in $t$. The energy per hole falls steeply as
holes are added [Fig.~4(b) of the main text]. With $k=4$ the holes form
two walls: the occupation of $A$ is depleted in the middle in favor of
$B$ and $C$ [Fig.~\ref{fig:dw_binding}], and the energy per hole comes
within $0.0006J_z$ of the isolated-wall value. With $k=6$ they form three
walls. The bond energies confirm this assignment. The four-hole state
lowers the $xy$ energy equally on the $\bm a_1$ and $\bm a_2-\bm a_1$
bonds and not on the circumferential $\bm a_2$ bonds, as expected for
one wall of each chirality, and the six-hole state, whose three walls
have the same chirality, lowers it on $\bm a_1$ only. A compact $\Psi$
region would use all three directions. The holes thus assemble into
walls although the cylinder does not require them.

\begin{figure}[!t]
\centering
\includegraphics[width=\columnwidth]{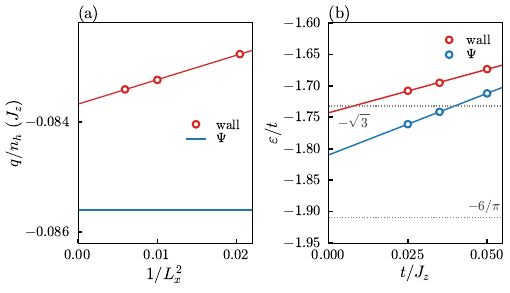}
\caption{Isolated wall and uniform $\Psi$ state ($n_h/N=1/9$) on the $W=6$
cylinder. (a) Energy per hole of wall arrays with spacing $L_x$ against
$1/L_x^2$ at $J_{xy}/J_z=0.1$, with the linear extrapolation to an isolated
wall, and the $\Psi$ state. (b) Energies per hole in units of $t$ against
$t/J_z$ ($J_{xy}/J_z=0.05$, $0.07$, $0.1$), with linear fits. Dotted lines:
the kink chain with $n_{\rm w}=2/3$ on a six-site ring ($-\sqrt3$) and the
untilted value $-6/\pi$.}
\label{fig:dw_wall}
\end{figure}

\begin{table}[!t]
\caption{Isolated wall and uniform $\Psi$ state ($n_h/N=1/9$) on YC
cylinders at $J_{xy}/J_z=0.1$. Energies are $q/n_h$ in units of
$J_z$; $\Delta=\varepsilon_{\rm wall}-\varepsilon_\Psi$. Extrapolated in
the truncation error, the $W=9$ values become
$\varepsilon_\Psi\approx-0.0856$ and $\Delta\approx+0.0040$.
$|\bn|$ is the site-summed bond-nematic amplitude per unit
wall length in the largest cell ($L_x=13$ for $W=6$, $11$ for $W=9$).}
\label{tab:dw_dmrg}
\begin{ruledtabular}
\begin{tabular}{lccc}
 & $W=6$ & \multicolumn{2}{c}{$W=9$}\\
 & $\chi=400$ & $\chi=800$ & $\chi=1600$\\
\hline
$\varepsilon_{\rm wall}$ & $-0.0837$ & $-0.0817$ & $-0.0817$\\
$\varepsilon_\Psi$ & $-0.0856$ & $-0.0847$ & $-0.0852$\\
$\Delta$ & $+0.0019$ & $+0.0031$ & $+0.0035$\\
$|\bn|$ per length & $0.61$ & $0.59$ & $0.59$\\
$\rho_s$ & $0.0039$ & $0.0034$ & $0.0035$\\
$(\pi/2)\rho_s$ & $0.0061$ & $0.0053$ & $0.0055$\\
\end{tabular}
\end{ruledtabular}
\end{table}

\subsection{Walls versus the \texorpdfstring{$\Psi$}{Psi} state}
Table~\ref{tab:dw_dmrg} and Fig.~\ref{fig:dw_wall}(a) compare an isolated wall with the uniform
$\Psi$ state at hole density $n_h/N=1/9$, which has the lowest energy
per hole among the uniform densities computed ($n_h/N=0.074$--$0.17$). On both circumferences $\Psi$ lies slightly below
the wall, by $\Delta=\varepsilon_{\rm wall}-\varepsilon_\Psi$, and $\Delta$ grows with $\chi$ because the gapless $\Psi$
state converges from above.

The wall energy agrees with the kink chain of
Eq.~\eqref{eq:wall_xx_chain} once the geometry is taken into
account. A wall that wraps a YC cylinder must run along the
circumference, which is tilted by $\pi/6$ from the direction of
equal $L$ and $R$ steps; this fixes
$n_{\rm w}=2/3$, and the zero-temperature limit of Eq.~\eqref{eq:wall_free_energy}
gives the kink gain $(6t/\pi)\sin(\pi n_{\rm w})$ per hole, i.e.,
$(6t/\pi)\sin(2\pi/3)$ instead of $6t/\pi$. For finite $W$, a
periodic chain of $W$ sites with $2W/3$ particles, optimized over
the boundary phase (set by the momentum of the wall along the
cylinder), gives $\sqrt3\,t$ per hole for $W=6$ and $1.69t$ for
$W=9$. iDMRG at $J_{xy}/J_z=0.05$, $0.07$, and $0.1$ on $W=6$ gives
$\varepsilon_{\rm wall}\approx-1.74t+1.4t^2/J_z$ [Fig.~\ref{fig:dw_wall}(b)], whose linear term
agrees with $\sqrt3\,t$ to within $1\%$, and the $W=9$ value
(Table~\ref{tab:dw_dmrg}) follows with the same $t^2$ term. The deficit of the linear-order gain
relative to $6t/\pi$ is therefore geometric; the rest of the deficit
at $J_{xy}/J_z=0.1$ is the $t^2/J_z$ correction.

\subsection{Bond fingerprint}
In every wall the $xy$ bond energy is lowered on a single bond
direction, the one along which the boson moves in the
$LR\leftrightarrow RL$ process of Eq.~\eqref{eq:wall_flip}. This bond
is set by the type of the wall, not by its orientation. On the
cylinder the wall runs along the circumference, so this bond makes
an angle $\pi/3$ with it, and $A\!\to\!B$ and $A\!\to\!C$ walls, which
have the same orientation there, resonate on the two different bond
directions that do not run around the circumference. For a wall with
$n_{\rm w}=1/2$ the bond is perpendicular to the wall. The map in
Fig.~4(a) of the main text uses a $W=9$, $L_x=19$ cell in which weak
fields on two columns of each domain, away from the wall, fix its
position; they raise the wall energy by less than $0.001J_z$ per hole.
Without them the wall delocalizes along the cell, with the same
single-direction bond pattern. The contribution to $|\bn|$ per unit wall length
(Table~\ref{tab:dw_dmrg}) depends only weakly on the wall spacing. In
the $\Psi$ state the three bond directions are equivalent on $W=9$; on
$W=6$ the bonds around the circumference are somewhat stronger, a
finite-width effect.

\subsection{Dependence on \texorpdfstring{$J_{xy}$}{Jxy}}
Figure~\ref{fig:dw_wall}(b) shows the wall and $\Psi$ energies per hole
on $W=6$ for $J_{xy}/J_z=0.05$, $0.07$, and $0.1$. In units of $t$ both
stay nearly constant while $t$ changes by a factor of two, so both gains
are of first order in $t$; the small linear drift is the $t^2/J_z$
correction. For the wall this reflects the first-order kink motion. For
$\Psi$, the bosons resonate on the honeycomb formed by two sublattices,
also at first order in $t$. The two states therefore compete on the same
scale $t$, with a difference $\Delta/t$ of a few percent over this range,
and the competition does not depend qualitatively on $J_{xy}$.


%


\bibliography{references}